\documentclass[prc,twocolumn,aps,showpacs,superscriptaddress]{revtex4}
\usepackage{amsmath}
\usepackage{amsfonts}
\usepackage{graphicx}
\begin{document}

\title{High-lying single-particle modes, chaos, correlational
entropy, and doubling phase transition}

\author{Chavdar Stoyanov}

\affiliation{Institute for Nuclear Research and Nuclear Energy,
Tzarigradsko Chaussee 72, 1784 Sofia, Bulgaria}

\author{Vladimir Zelevinsky}

\affiliation{National Superconducting Cyclotron Laboratory and
Department of Physics and Astronomy, Michigan State University,
East Lansing, Michigan 48824-1321}

\begin{abstract}

Highly-excited single-particle states in nuclei are coupled with
the excitations of a more complex character, first of all with
collective phonon-like modes of the core. In the framework of the
quasiparticle-phonon model we consider the structure of resulting
complex configurations using the $1k_{17/2}$ orbital in $^{209}$Pb
as an example. Although, on the level of one- and two-phonon
admixtures, the fully chaotic GOE regime is not reached, the
eigenstates of the model carry significant degree of complexity
that can be quantified with the aid of correlational invariant
entropy. With artificially enhanced particle-core coupling, the
system undergoes the doubling phase transition with the
quasiparticle strength concentrated in two repelling peaks. This
phase transition is clearly detected by correlational entropy.

\end{abstract}


\date{\today}

\maketitle


\section{Introduction}

It is now firmly established that many-body quantum dynamics
acquire typical features of quantum chaos in the region of high
level density. The interrelation between many-body physics and
quantum chaos was studied, in particular, in the framework of the
nuclear shell model with realistic and random interactions, see
\cite{brody,big,kota,meso} and references therein. A description
of a stable many-body system starts with the mean field and
quasiparticles determined by the symmetry of the mean field. As
excitation energy and level density increase, residual
interactions convert stationary states of independent
quasiparticles into exceedingly complex superpositions of many
basis states. The many-body structure at this stage should be
described in statistical terms, and in the limiting case of
extreme chaos it locally approaches the predictions of the
Gaussian Orthogonal Ensemble (GOE) of random matrices.

The complexity of stationary many-body states can be quantified
with the aid of such characteristics as information entropy or
inverse participation ratio \cite{izr,entr,big,kota} of generic
wave functions. These quantities smoothly change with excitation
energy revealing strong mixing of original states and loss of
simple quantum numbers, typical features of quantum chaos. However
such measures of complexity are unavoidably associated with a
starting basis, in these cases determined by the mean field. They
actually provide the localization length of a complex state in a
chosen basis with no information on the correlations still present
in the wave function.

Different characteristics can be used in order to probe the
sensitivity of the system to external perturbations. In classical
case such sensitivity is known to be the main property of
dynamical chaos. A special entropy-like quantity, called {\sl
invariant correlational entropy} (ICE), was suggested in Ref.
\cite{ICE} as a measure of complexity related to the response of a
system to a random noise included in the Hamiltonian. This
quantity is by construction invariant with respect to the basis
transformations. It also reflects the correlations and phase
relationships between the components of the wave functions as far
as they are revealed in the response of the system to a
perturbation. Taking the strength of the interaction as a control
parameter, one can clearly see the quantum phase transitions as it
was demonstrated in the interacting boson model \cite{cej} and in
the evolution of pairing in the shell model \cite{VZ}.

The problem of the interaction of a quasiparticle with an
even-even core is of considerable interest for nuclear physics. It
was quite well studied for low-lying excited states where the
interactions of the quasiparticle predominantly with quadrupole
and octupole surface vibrations are important. At higher
excitation energy, a large number of other nuclear modes influence
the damping of the quasiparticle motion. The mixing of the simple
mode with the states of the next levels of complexity leads to the
fragmentation of the single-particle strength over a wide domain
of excitation energy, $-$ the single-particle state obtains a
spreading width.

It was shown in Ref. \cite{Gal} that the spreading occurs in two
stages. At the first stage of fragmentation, the single-particle
state is spread over several doorway states. At the second step,
the doorway states are spread through the mixing with many complex
excitations related to other degrees of freedom. The limiting case
of self-consistent hierarchy of multi-step spreading was discussed
in Ref. \cite{LBBZ}.

The highly-excited single-particle mode has been experimentally
studied via one-nucleon transfer reactions \cite{Gal}. For
example, broad structures located around 10 MeV excitation energy
have been observed in the reaction $^{208}$Pb$(\alpha,^3$He)
$^{209}$Pb. It was shown \cite{Gal94} that the structures are
connected with the excitation of high-spin orbitals $1j_{13/2},
2h_{11/2}$ and $1k_{17/2}$. The fragmented wave functions based on
these states contain many components. The related complexity of
the wave function was analyzed in Ref. \cite{Sol95}. It was
concluded that the damping of a simple mode cannot be
unambiguously associated with the random contribution of more
complex components to the structure of the states.

The goal of the presented study is to introduce a new measure of
the complex structure of highly-excited states. The paper is
organized as follows. In Sec. II the formalism of correlation
entropy is introduced. In Sec. III  the main features of the used
model for the structure of excited states are presented. The
complexity of of the excited states and the connection to the ICE
is discussed in Sec. IV. The doubling phase transition in the
regime of strong coupling in odd nuclei is demonstrated in Sec. V.
Finally, in Sec. VI the conclusions are drawn.

\section{Correlational entropy}

To study the sensitivity of the excited states to variation of
external parameters we use the invariant correlational entropy
(ICE) \cite{ICE}. The ICE method presumes that Hamiltonian
$H(\lambda)$ of a system depends on a random parameter $\lambda$.
The parameter $\lambda$ (``noise") is considered as a member of an
ensemble characterized by the normalized distribution function
${\cal{P}}(\lambda)$,
\begin{equation}
\int d{\lambda}{\cal{P}}(\lambda) =1.            \label{1}
\end{equation}
In an arbitrary primary basis $|k \rangle$, we follow the
evolution of any stationary state $|\alpha;\lambda\rangle$ as a
function of $\lambda$. At a given value of $\lambda$, the state
can be decomposed as
\begin{equation}
|\alpha;\lambda \rangle = \sum_k C^{\alpha}_k(\lambda)|k\rangle.
                                               \label{2}
\end{equation}

The ICE is  defined as
\begin{equation}
S^{\alpha} =-{\rm Tr}\{\varrho^{\alpha} \ln (\varrho^{\alpha})\},
                                              \label{3}
\end{equation}
where $\varrho^{\alpha}$ is the density matrix of the state
$|\alpha \rangle$ averaged over the noise ensemble. In the basis
$|k\rangle$ prior to the averaging we construct first this matrix
for a given value of $\lambda$ as
\begin{equation}
\varrho_{kk^{\prime}}^{\alpha}(\lambda) = C_k^{\alpha}(\lambda)
C_{k^{\prime}}^{\alpha*}(\lambda),              \label{4}
\end{equation}
and then average over the ensemble,
\begin{equation}
\varrho_{kk^{\prime}}^{\alpha}= \int d\lambda\, {\cal P}(\lambda)
\varrho_{kk^{\prime}}^{\alpha}(\lambda).         \label{5}
\end{equation}
While the density matrix $\varrho^{\alpha}$, eq. (\ref{4}), of a
pure state, being a projector onto the state $|\alpha\rangle$ and
having correspondingly only one nonzero eigenvalue equal to one,
leads to zero entropy $S^{\alpha}$, the averaged density matrix
(\ref{5}) has its eigenvalues between 0 and 1 and produces
non-zero correlational entropy.

In contrast to {\sl information} (Shannon) entropy $I^{\alpha}$ of
the same state $|\alpha\rangle$, conventionally used for
quantifying the complexity,
\begin{equation}
I^{\alpha}=-\sum_{k}|C^{(\alpha)}_{k}|^{2}\ln(|C^{(\alpha)}_{k}|^{2}),
                                                    \label{6}
\end{equation}
the ICE is {\sl basis-independent} von Neumann entropy that
reflects the correlations between the wave function components
which are subject to fluctuations determined by the parameter
$\lambda$. The value $S^{\alpha}$ for a given state typically
increases with the complexity of the state and reaches the maximum
at the point where the change of the parameter around some average
value implies the most radical change of the structure of the
system. Such a point (in fact, a region in finite systems) can be
identified with the quantum phase transition or crossover, and in
the vicinity of this point the structural fluctuations of the wave
functions are strongly enhanced. At further change of the average
value of the noise parameter, the ICE, as a rule, goes down. This
means that the stronger interaction has established a new order
with greater rigidity with respect to fluctuations of parameters.
In this way the sharp boundaries between the different symmetry
classes very confirmed \cite{cej} in the interacting boson models,
and the critical strengths of isovector and isoscalar pairing in
the $sd$ shell model were established \cite{VZ}.

\section{The model of excited states in odd nuclei}

We adopted the {\sl Quasiparticle-Phonon Model} (QPM) by Soloviev
at al. \cite{Sol} to describe the properties of highly-excited
states in odd spherical nuclei. According to the model, the
Hamiltonian of the system of an odd number $A+1$ particles has the
form
\begin{equation}
H = h + H_{{\rm core}} + H_{{\rm coupl}}.       \label{7}
\end{equation}
The first term, $h$, describes the motion of a quasiparticle in a
mean field potential $U$ created by the even-even core,
\begin{equation}
h = -\frac{1}{2m}\,\nabla^2 + U~.                \label{8}
\end{equation}
The core Hamiltonian is a sum of single-particle Hamiltonians
$h_i$ and two-body residual interactions $V_{i,j}$,
\begin{equation}
H_{{\rm core}} = \sum_{i=1}^A h_i +\sum_{i<j}^A V_{i,j}~.
                                                  \label{9}
\end{equation}
The last term in eq. (\ref{7}) is a sum of interactions between
the odd quasiparticle and particles in the core,
\begin{equation}
H_{{\rm coupl}} = \sum_{i=1}^A V_{0,i}~.        \label{10}
\end{equation}
In the spirit of the QPM, the Hamiltonian $H_{{\rm core}}$ is
treated in the random phase approximation (RPA), i.e. the
particle-hole configurations are built with the subsequent RPA
diagonalization.

The properties of the (A+1)-nucleus can be described in terms of
the quasiparticle states $\alpha^\dagger_a\mid 0 \rangle$ ,
quasiparticle-plus-phonon states $[\alpha^\dagger_a\otimes
Q^\dagger_\nu] \mid 0 \rangle$ and quasiparticle-plus-two-phonon
states $[\alpha^\dagger_a\otimes [Q^\dagger_\mu\otimes
Q^\dagger_\nu]] \mid 0 \rangle$, where all combinations have the
same total spin and parity quantum numbers $J^{\pi}M$. Here
$\alpha _{a}^{\dagger}$ is the quasiparticle creation operator
with shell-model quantum numbers $a\equiv (n,l,j,m)$, whereas
$Q^{\dagger}_{\nu}\equiv Q_{\lambda \mu i}^{\dagger}$ denotes the
phonon creation operator with the angular momentum $\lambda$,
projection $\mu $ and RPA root number $i$.

The following wave functions describe in the QPM the ground and
excited states of the odd nucleus with angular momentum $J$ and
projection $M$:
\[ \Psi ^\alpha _{JM}= C_J^\alpha\{ \alpha_{JM}^\dagger+
\sum_{a\nu}D_a^{\nu}(J\alpha )\left[ \alpha _{a}^\dagger
Q_{\nu}^\dagger \right] _{JM}\]
\begin{equation}
+\sum_{a\nu\nu'}F_{a}^{\nu\nu'}(I;J\alpha )\left[ \alpha
_{a}^\dagger \left[Q_{\nu}^\dagger Q_{\nu'}^\dagger \right]_{I}
\right] _{JM}\} \Psi _{0}~.                         \label{11}
\end{equation}
The wave function  $\Psi_{0}$ represents the ground state of the
neighboring even-even nucleus (also quasiparticle and phonon
vacuum) and $\alpha$ stands for the number within a sequence of
states of given $J^\pi$. The coefficients $C_J^\alpha$,
$D_a^{\nu}(J\alpha)$ and $F_{a}^{\nu\nu'}(I;J\alpha)$ are the
quasiparticle, quasiparticle + phonon and quasiparticle + two
phonons amplitudes, respectively, for the state $\alpha$. The norm
of the wave function (\ref{11}) reads:
\[\langle\Psi^{\alpha\ast}_{JM}|\Psi^{\alpha}_{JM}\rangle\]
\begin{equation}
=(C_J^\alpha)^2\left\{1+\sum_{a\nu}[D_a^{\nu}(J\alpha)]^2+
2\sum_{a\nu\nu'}[F_{a}^{\nu\nu'}(I;J\alpha)]^2\right\}. \label{12}
\end{equation}

The Hamiltonian (\ref{7}) contains several parameters. The mean
field (\ref{8}) was chosen to be the Woods-Saxon potential. The
residual interaction (\ref{9}) in the particle-hole channel was
taken in a separable form with interaction strengths in each mode
considered as adjusted parameters. The quasiparticle-phonon
coupling term (\ref{10}) does not contain any additional freedom.
The core excitations with all spins and natural parity, $1^-, 2^+,
3^-, 4^+, 5^-, 6^+, 7^-$ and $8^+$, were included in the
calculations. For each momentum and parity, the RPA states up to
20 MeV excitation energy were taken into account. The large phonon
basis is necessary for correct description of the single-particle
strength distribution in a broad range of excitation energy.

The used set of parameters have been successfully applied to
describe the properties of low-lying as well as highly-excited
states in $^{209}$Pb, see for example Ref. \cite{Giai}. For
studying the ICE, we selected the single-particle $1k_{17/2}$
state. The properties of this state are studied in detail in Ref.
\cite{Giai}. This orbital is quasi-bound in the Woods-Saxon
potential being located at 4.88 MeV, energy much higher than the
Fermi level of $^{209}$Pb. Because of its high energy, the state
is surrounded by many quasiparticle-plus-phonon and
quasiparticle-plus-two-phonon states.

\section{Complexity of states and correlational entropy}

At high excitation energy, the level density is large and it is
convenient to calculate the single-particle strength distribution
by means of the strength function \cite{Gal} using an averaging
Lorentzian function. The distribution of the single-particle
strength [coefficient $C^2$ of eq. (\ref{11})] of the
$1k_{17/2}$-state is shown in Fig. 1. The number of
quasiparticle-plus-phonon components included in the wave function
(\ref{11}) is 420, while the number of
quasiparticle-plus-two-phonon components is 1116. The Lorentzian
smoothing parameter was chosen to be 0.2 MeV.

\begin{figure}
\begin{center}
\includegraphics[scale=.33,angle=270]{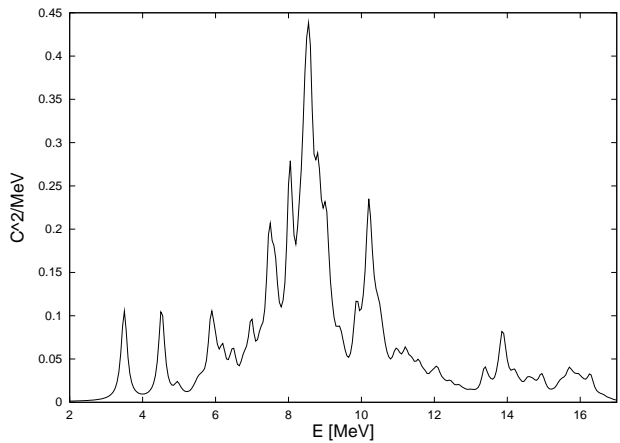}
\end{center}
\begin{center}
\caption{\label{Fig. 1}
Distribution of the single-particle strength of \\
the state $1k_{17/2}$ in $^{209}$Pb.}
\end{center}
\end{figure}

It is seen from Fig. 1 that the single-particle strength is spread
over a broad interval of excitation energy. The largest fraction,
81 \% of the strength, is concentrated between 5 and 13 MeV
excitation energy, and the state acquires the large spreading
width in this domain, $\Gamma^{\downarrow}=1.5\,$ MeV
($\Gamma^{\downarrow}$ is calculated as the second moment of the
distribution). It was shown \cite{Gal,Ber} that the main mechanism
leading to the spreading is the interaction of the particle with
the vibrations of the even core.

In the case of $^{209}$Pb the lowest core excitation is  $3^-_1$
state. The component $[1j_{15/2} \otimes 3^-_1]_{17/2^{+}} $
dominates in the structure of the lowest two excited states, where
$8\,\%$ of the single-particle strength is concentrated. The
component $[2g_{9/2} \otimes 4^+_1]_{17/2^{+}}$ also contributes
to the structure of these excited states. The contribution of the
components, where the particle is coupled to $2^+_1,\,4^+_1$ and
$6^+_1$ phonons, is more important at the higher part of the
distribution (above 13 MeV). The coupling of particle with $5^-$
phonons is significant in the domain of the main peak \cite{Giai}.
The $7^-$ and $8^+$ phonons mainly contribute at excitations
around 10 MeV \cite{Giai}. More complex
quasiparticle-plus-two-phonon components influence predominantly
the secondary fragmentation of quasiparticle-plus-one-phonon
components.

The correlation entropy (\ref{3}) connected with the excited
states was calculated using two types of the distribution
function, the Lorentzian function,
\begin{equation}
{\cal{P}}(\lambda) = \frac{1}{2 \pi} \frac{\Delta} {\lambda^2 +
\frac{\Delta^2}{4}},                         \label{13}
\end{equation}
and normal (Gaussian) distribution,
\begin{equation}
{\cal{P}}(\lambda) = \frac{1}{\sigma \sqrt{2 \pi}} \exp{
\frac{-{\lambda}^2} { 2 {\sigma^2}}}.    \label{14}
\end{equation}
The value of the parameters used in the calculations was
$\Delta=0.8$ and $\sigma=0.5$. For this choice of the parameters
both distributions have the same maximum value.

The calculation was done as follows. For the values of
$\lambda=\lambda_c$ the matrix elements of $H_{{\rm coupl}}$, eq.
(\ref{10}), are multiplied by the corresponding value of
${\cal{P}}(\lambda_c)$. The new matrix elements are used to
calculate the eigenvalues and eigenvectors of the Hamiltonian
(\ref{7}). The elements of the density matrix
${\cal{\varrho}}^{\alpha}_{kk^{\prime}} (\lambda_c)$ are
calculated according to the eq. (\ref{4}) and averaged, eq.
(\ref{5}). In the case of the Lorentzian distribution, the value
of $\lambda_c$ is changed in  the interval $[-3\Delta, 3\Delta]$,
while in the case of the normal distribution function $\lambda_c$
is taken from the interval $[-2.5\sigma, 2.5\sigma]$.

To estimate quantitatively the absolute value of the ICE we use
also the summed correlational entropy of $N$ states,
\begin{equation}
F(N) = \sum_{\alpha=1}^N S^{\alpha},              \label{15}
\end{equation}
where $\alpha$ labels the eigenstates in the sector with given
quantum numbers. The normalized value of $F(N)$ is defined as
\begin{equation}
F=\frac{1}{N}F(N).                                 \label{16}
\end{equation}

The wave function (\ref{11}) includes many components. Because of
the different structure of the components, the values of the
matrix elements of $H_{{\rm coupl}}$ reveal large fluctuations.
There are many small matrix elements and the corresponding
components are weakly correlated. Their contribution to the
spreading process has to be negligible. The ICE gives the
opportunity to construct the appropriate basis including only the
most important components. The dependence of the ICE on the
dimension of the basis is presented in Fig. 2. The basis is
truncated according to the criterion connected with the maximum
matrix element. All components having the matrix element of the
coupling with the original single-particle state less than a
certain fraction of the maximum matrix element are not included in
the calculation. Here only the one-phonon components
$[\alpha^\dagger_a\otimes Q^\dagger_\nu] \mid 0 \rangle$ of the
wave function (\ref{11}) are taken into account. It is seen that
the normalized entropy, eq. (\ref{16}), reveals saturation for the
small values of the truncation parameter. The components, whose
matrix element is less than $5 \%$ of the maximum one, contribute
weakly to $F$. The number of such components is large but their
influence on the spreading process is negligible.

\begin{figure}
\begin{center}
\includegraphics[scale=.33,angle=270]{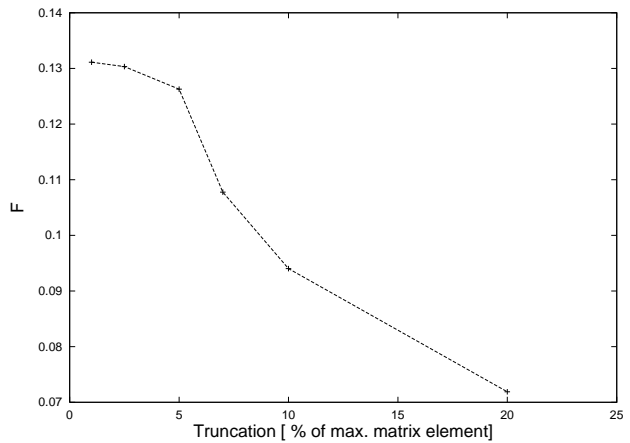}
\end{center}
\caption{\label{Fig. 2} Dependence of correlation entropy on the
truncation of the quasiparticle-plus-phonon basis.}
\end{figure}

The results for the ICE found with the two distribution functions
(\ref{13}) and (\ref{14}) are compared in Fig. 3. Fig. 3$a$
displays the correlation entropy of the obtained states in the
case of the normal distribution, eq. (\ref{14}). There is a
background of small values of the ICE due to the large amount of
weakly interacting states. The greater values of entropy are
located in the vicinity of the peaks of the single-particle
strength distribution pointing to the regions where the proximity
of many strongly coupled states leads to enhanced sensitivity of
the wave functions. As mentioned above, in the region around 8 MeV
the particle is coupled predominantly with the $3^-_1$ phonon,
while around 13 MeV the coupling is occurring mainly with the
first positive parity phonons. One can see the tendency to the
enlargement of entropy when the excitation energy increases and a
larger density of the states at higher excitation energy enters
the game.

The calculation of the ICE with the Lorentzian distribution
function shown in Fig. 3$b$ leads to the results similar to those
in Fig. 3$a$. The regions including the enhanced density of states
with large entropy appear at the same excitation energy. In the
case of the normal distribution the value of $F(N)$, eq.
(\ref{15}), is 37.50, while for the Lorentzian it is 54.44, mainly
because the Lorentzian has the long tails. In spite of this
difference in the value of $F(N)$, the main features of both
distributions of the ICE are similar being determined mostly by
the interaction and density of states and only marginally by the
type of the noise distribution function.

\begin{figure}
\begin{center}
\includegraphics[scale=.33,angle=270]{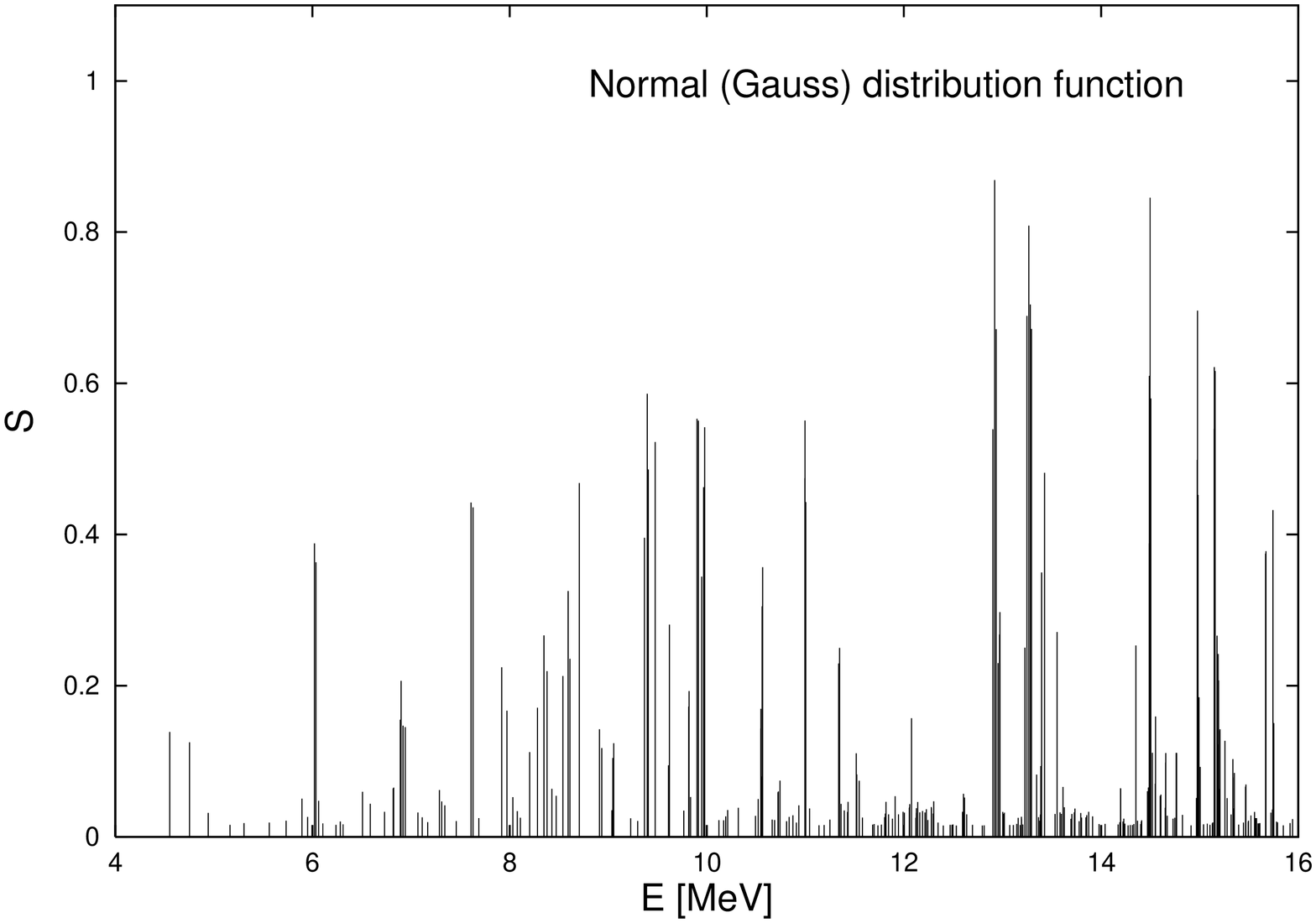}
\end{center}
\begin{center}
\includegraphics[scale=.33,angle=270]{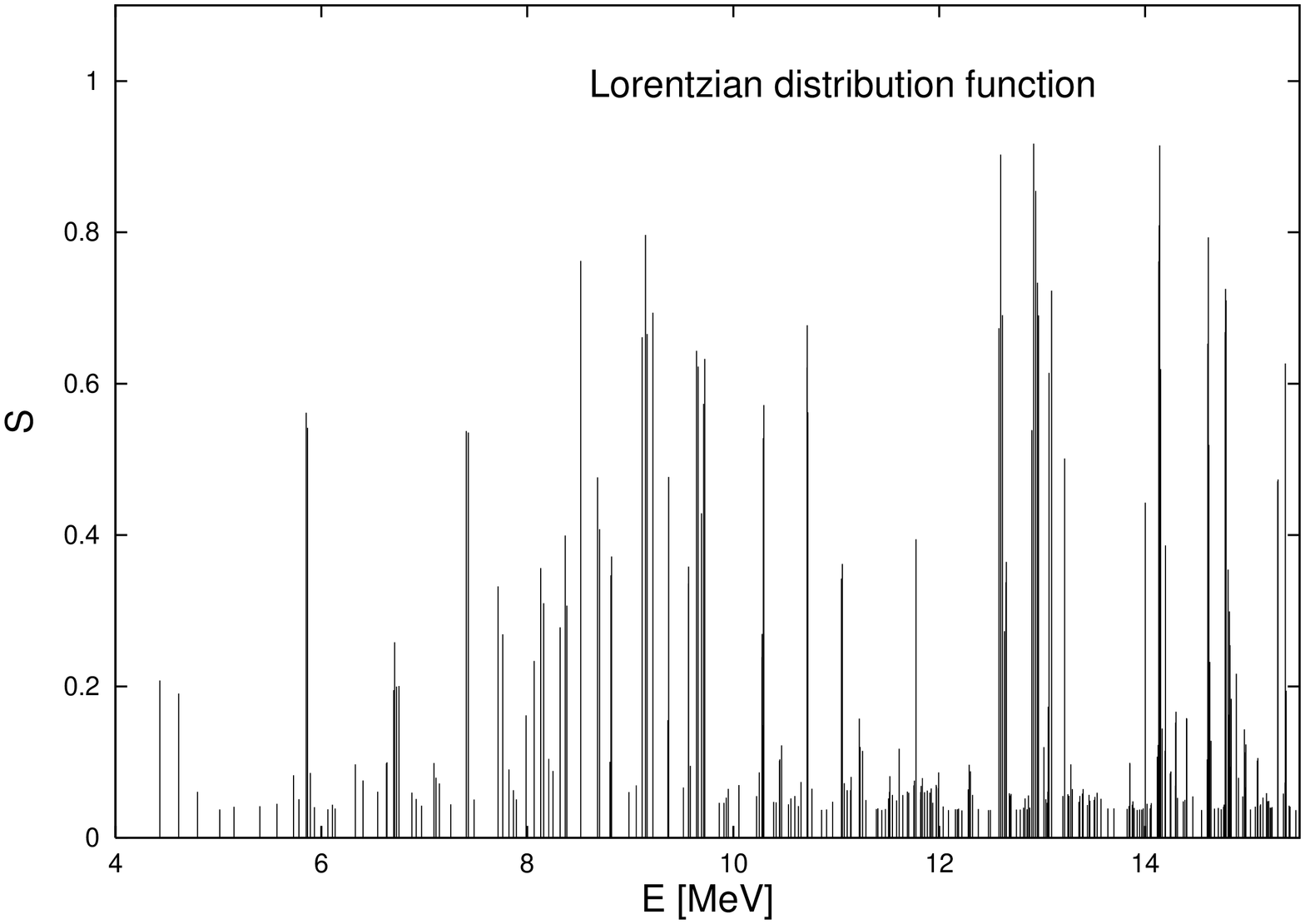}
\end{center}
\caption{\label{Fig. 3} Correlation entropy of the excited states
${17/2}^+$ in $^{209}$Pb. a) The normal (Gaussian) distribution
function is used in the calculation for upper panel, and b) the
Lorentzian distribution is used for lower panel. The wave function
includes only quasiparticle-plus-phonon components.}
\end{figure}

To test the sensitivity of the results to the parameters of the
model, we introduce a new parameter $k$. Multiplying the matrix
elements of $H_{{\rm coupl}}$, eq. (\ref{10}), by this number we
vary the overall strength of the particle-core coupling. The value
of $k=1$ corresponds to the realistic strength of $H_{{\rm
coupl}}$ chosen according to the initially fitted parameters. This
case is presented in Fig. 3. Only the components
$[\alpha^\dagger_a\otimes Q^\dagger_\nu] \mid 0 \rangle$ of the
wave function (\ref{11}) are taken into account.

The absolute value of entropy $F(N)$, eq. (\ref{15}), depends on
the value of $k$, i.e. $F(N) \Rightarrow F(N,k)$; the function
$F(N,k)$ is shown in Fig. 4 revealing a pronounced maximum at
$k_{c}=1.6$. The maximum of $F(N,k)$ at $k=k_{c}$ can be
identified with a quantum {\sl phase transition} smeared as
expected in a finite system. The single-particle state interacts
strongly with the sea of more complex excitations, and its
contribution to the structure of any individual excited state is
strongly reduced. But it has to be pointed out that even for the
case of maximum entropy, $k=1.6$, the single-particle component
preserves its dominance in the main peaks of the strength
distribution. When the more complex quasiparticle-plus-two-phonon
components are included in the wave function (\ref{11}), the
correlation entropy rapidly increases. The value of normalized
entropy (eq. \ref{16}) for the case presented in Fig. 1 is larger
than the maximum value shown in Fig. 2 by a factor close to 6. The
additional correlations are mainly due to the coupling of
qusiparticle-plus-one- and quasiparticle-plus-two-phonon
components. Because of this hierarchy of couplings, the
distribution of the single-particle strength is not affected too
much by the new terms in the wave function (\ref{11}).

\begin{figure}
\begin{center}
\includegraphics[scale=.33,angle=270]{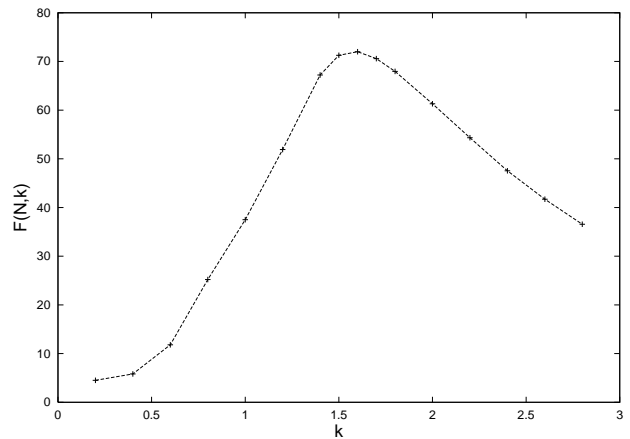}
\end{center}
\caption{\label{Fig. 4}
 Dependence of the integral correlation entropy on the
strength of particle-core coupling. Only the components
$[\alpha^\dagger_a\otimes Q^\dagger_\nu] \mid 0 \rangle$ of the
wave function (\ref{11}) are taken into account.}
\end{figure}

For small values of $k$, the mixing of states is suppressed and
the ICE decreases rapidly, see Fig. 4. This case corresponds to
the weak particle-core coupling and relatively simple wave
function. The strength of single-particle states is distributed in
a narrow vicinity of its unperturbed energy. The nearest-neighbor
level spacing distribution for $k=0.2$ is very close to the
Poisson distribution. In the case when only
quasiparticle-plus-phonon components are included in the wave
function (\ref{11}), even at $k=1$ the correlations are quite
large, Fig. 4. This influences the level spacing distribution of
$17/2^+$ states. The nearest-neighbor spacing distribution of
$17/2^+$ excitations can be fit by the Brody distribution
\cite{brody} with the Brody parameter equal to 0.4. At this stage
the single-particle component is not completely smeared and
chaotically distributed over the excitations of the system. Clear
remnants of the simple excitation seen in the complicated wave
functions are similar to the phenomenon of scars \cite{scar}
existing in simple quantum systems.

The nearest-neighbor spacing distribution of $17/2^+$ excitations
for the case when the quasiparticle-plus-two-phonons components
are included in the wave function (\ref{11}) is shown in Fig. 5.
The distribution is calculated for the region of the main
single-particle strength, 5-13 MeV, where the density of the
states is not changed much. As seen from Fig. 5, the distribution
can be described by the Brody distribution with the parameter
equal to 0.6. The higher degree of chaoticity indicates the
importance of more complex components and their influence on the
damping process. At the first stage, when the single-particle
states interact only with quasiparticle-plus-one-phonon
components, the regularities induced by the mean field, such as
the structure of the level density, are not completely destroyed.
At the next stage, when the interaction with the components of the
next level of complexity is switched on, these regularities are
partly smeared, but the wave function (\ref{11}) is still
relatively simple. One can recall old results \cite{BZ}, where the
coupling of an unpaired particle with the collective monopole and
quadrupole modes was considered in detail (later these results
were applied \cite{Urin}) to the spreading width of giant
resonances due to their mixing with low-lying shape vibrations).
In exactly solvable models with a particle attached to a single
level it was shown \cite{BZ} that the main effect of the
particle-phonon coupling is in creating a coherent state of the
phonon field. The chaotic elements should be associated with the
mixing between various quasiparticle levels and coupling with
different phonon modes.

\begin{figure}
\begin{center}
\includegraphics[scale=.33,angle=270]{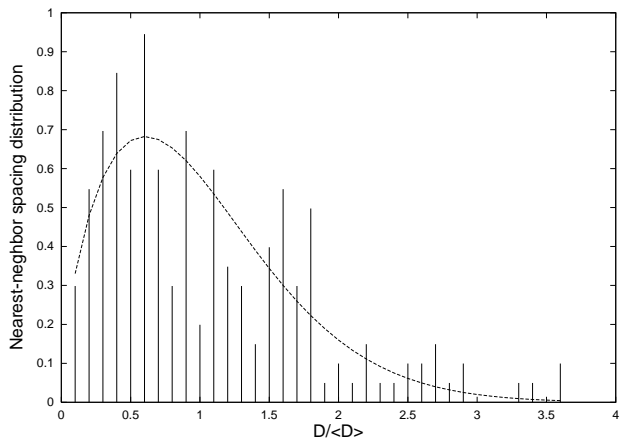}
\end{center}
\caption{\label{Fig. 5}
 Nearest-neighbor spacing distribution of $17/2^+$ states
in $^{209}$Pb. The calculated values are fit by the Brody
distribution with the Brody parameter equal to 0.6.}
\end{figure}

More multi-phonon components are to be taken into account for
reaching the full complexity. As degree of complexity grows, the
wave function (\ref{11}) approaches the realistic wave function
necessary to describe the complex stricture of high-lying excited
states. It has to be pointed out that a few quasiparticle
components are weakly influenced by the growth of complexity. By
this reason, even a rather simple wave function, truncated for
example, by the quasiparticle and quasiparticle-plus-phonon
components can be used to describe the transition probabilities
connecting high-lying and low-lying excited states. For this
purpose, a simple practical procedure of averaging high-lying
single-particle strength can be used \cite{Tson}.

\section{Doubling phase transition}

For large values of $k$ the particle-core coupling becomes very
strong. The single-particle strength is split into two main pieces
repelled to low and high excitation energies. The case for $k=2.0$
is shown in shown in Fig. 6. The single-particle strength
corresponding to the low and high energy peaks is $31\,\%$ and
$21\,\%$, respectively. The rest of the strength is distributed in
between the peaks.

\begin{figure}
\begin{center}
\includegraphics[scale=.33,angle=270]{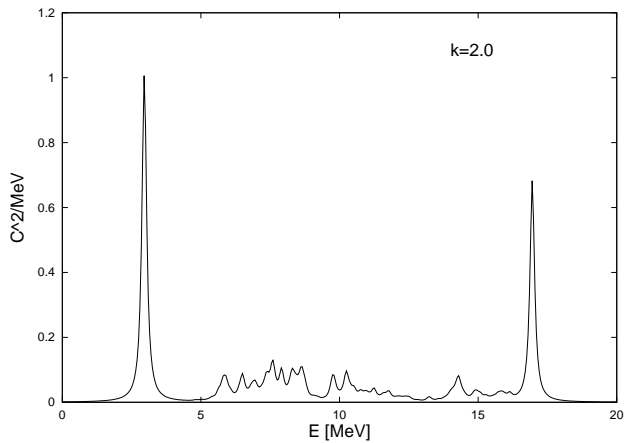}
\end{center}
\caption{\label{Fig. 6}
 Distribution of the single-particle strength of the state
$1k_{17/2}$ in $^{209}$Pb for large value of particle-core
coupling after the doubling phase transition. The value of $k$ is
$k=2.0$.}
\end{figure}

To identify the physical nature of the quantum phase transition
that occurs at an enhanced particle-phonon coupling, we can
analyze the behavior of the strength function of the original
single-particle state coupled with states of a more complex
character. Such an analysis was first presented in Ref. \cite{Lew}
in application to giant resonances. Our formulation is in fact a
particular limiting case of a generic problem \cite{BM} of a
``bright" state coupled to the background that contains a
hierarchy of states of increasing complexity. In this
consideration, the bright state $|0\rangle$, in our case the bare
single-particle state with unperturbed energy $E_{0}$, is mixed
with the background states $|\nu\rangle$, that contain, apart from
the quasiparticle, the phonon excitations of the core with
unperturbed energies $\epsilon_{\nu}$. The eigenstates
$|\alpha\rangle$ that emerge from this mixing are complicated
superpositions
\begin{equation}
|\alpha\rangle=C_{0}^{\alpha}|0\rangle
+\sum_{\nu}C_{\nu}^{\alpha}|\nu\rangle.         \label{17}
\end{equation}
Their energies $E_{\alpha}$ are the roots of the secular equation
\begin{equation}
X(E)\equiv E-E_{0}-\sum_{\nu}\frac{|V_{\nu}|^{2}}
{E-\epsilon_{\nu}}=0,                                \label{18}
\end{equation}
where $V_{\nu}$ are the matrix elements of coupling of the bright
state with the background states. The fragmentation of the
strength over the eigenstates $|\alpha\rangle$ is given by
\begin{equation}
|C_{0}^{\alpha}|^{2}=\left(\frac{dX}{dE}\right)_{E=E_{\alpha}}=
\left[1+\sum_{\nu}\frac{|V_{\nu}|^{2}}
{(E_{\alpha}-\epsilon_{\nu})^{2}}\right]^{-1}.   \label{19}
\end{equation}
The strength function of the bright state is found as
\begin{equation}
{\cal F}(E)=\sum_{\alpha}|C_{0}^{\alpha}|^{2}\delta(E-E_{\alpha}).
                                           \label{20}
\end{equation}

If the quantities $V^{2}$, the average value
$\langle|V_{\nu}|^{2}\rangle$ of coupling matrix elements squared,
and $D$, the mean level spacing in the background, can be
considered to be weakly fluctuating from one state $|\nu\rangle$
to another, the result of the mixing is determined by the ratio of
the ``standard" spreading width,
\begin{equation}
\Gamma_{s}=2\pi\,\frac{V^{2}}{D},              \label{21}
\end{equation}
to the energy range $a\simeq ND$, where $N$ is a number of
effectively interacting background states. Typically, one has in
average a Breit-Wigner strength function \cite{BM} with the width
(\ref{21}) for $\Gamma_{s}\ll a$; the further evolution as a
function of increasing coupling leads to the Gaussian shape with
the width increasing towards and beyond $a$ \cite{Lew,Fraz,FI,KS}.
In the transitional region the dependence of the strength function
upon the coupling intensity $V$ changes from the quadratic to
linear \cite{Lew,Fraz}. Finally, at even stronger coupling, eqs.
(\ref{18}) and (\ref{19}) predict a phase transition to the new
situation when the strength is accumulated in two peaks on both
sides of the centroid, see Fig. 2 in Ref. \cite{Lew}. In the
extreme limit, the peaks are pushed out of the region of the
background states,
\begin{equation}
E\approx E_{0}\pm \left[\sum_{\nu}|V_{\nu}|^{2}\right]^{1/2},
                                            \label{22}
\end{equation}
and the strength of the original state, according to eq.
(\ref{19}), is evenly divided between the peaks.

The physical mechanism of this {\sl doubling phase transition}
(known also in quantum optics) is the following. In the regime of
strong coupling, the background states of the same symmetry turn
out to be intensely interacting among themselves through the
bright state. Effectively this interaction is close to the
factorized one built as a product of the matrix elements to and
from the bright state. The factorized coupling leads to the
formation of the second collective state as a coherent
superposition of the background states. This state accumulates a
significant strength, and the repulsion between the two collective
states leads to the two-peak pattern. We see that the doubling
phase transition is adequately identified by the maximum of
correlational entropy.

\section{Conclusion}

In the presented study the recently suggested measure of the
complexity was tested for the wave function of a high-lying
excited quasiparticle state in a system where the fermionic
quasiparticle strongly interacts with bosonic collective
excitations of the core. The new quantity $-$ invariant
correlational entropy $-$ was used to estimate the growth of
complexity as a result of admixture of many new components to the
wave function of a quasiparticle.

In contrast to information entropy that displays the degree of
complexity of individual states with respect to a certain
reference basis, the ICE is basis-independent reflecting mostly
the sensitivity of a given state to the external noise. It was
shown that the ICE depends mainly on the interaction strength and
density of the background levels but less on the distribution
function used as a noise generator.

We have also calculated the neighboring level spacing distribution
as a conventional indicator of quantum chaos. It is shown that
this distribution is correlated with ICE moving in the direction
of the Wigner distribution characteristic for the Gaussian
orthogonal ensemble (but not reaching this limit) in parallel to
the increase of the value of ICE. Although the model wave function
truncated on the level of the quasiparticle-plus-two-phonon
components is not sufficiently chaotic to manifest entirely the
complicated structure of high-lying excited states, it can be used
to describe the distribution of a few quasiparticle components.
These components in fact play the role similar to the scars in
quantum chaos.

As a function of the overall strength of quasiparticle-phonon
interaction, the ICE increases and reaches a pronounced maximum at
the coupling strength equal to 1.6 of its realistic value. In this
region the system would undergo the quantum doubling transition
when the original single-particle excitation forms its own
counterpart built of more complicated states with the same quantum
numbers and the two peaks repel each other and share the original
strength. The ICE properly reflects this transformation. The
question remains whether such a phenomenon, known in quantum
optics, could be observed in real nuclear spectra.

Similar calculations have been also performed for the excited
neutron deep-hole orbital $1h_{11/2}$ in $^{207}$Pb and high-lying
proton orbital $1i_{13/2}$ in $^{145}$Eu. The obtained results are
in agreement with the above conclusions which confirms generic
features of underlying physics.\\
\\
The work was partly supported by the Bulgarian-American Commission
for Educational Exchange (Fulbright grant 03-2106) and Bulgarian
Science Foundation (Contract  Ph. 1311). Support from the NSF,
grant PHY-0244453, is highly appreciated.

\end{document}